\documentclass[journal,draftcls,onecolumn,12pt,twoside]{IEEEtranTCOM}
\newcommand{\floor}[1]{\lfloor #1 \rfloor}
\usepackage{graphicx, amssymb, cite, amsmath, amsfonts, epsfig}
\usepackage[]{algorithm2e}
\graphicspath{{./figures/}}
\normalsize
\ifCLASSINFOpdf
\else
\fi
\begin{document}
	\title{Symbolic Stochastic Chase Decoding of Reed-Solomon and BCH Codes}
	\author{Hossein~Mani,~\IEEEmembership{Student~Member,~IEEE}
		and~Saied~Hemati,~\IEEEmembership{Senior~Member,~IEEE}
		\thanks{The authors are with the Department of Electrical and Computer Engineering, University of Idaho, Moscow, ID, 83843 USA. Emails: \{hmani, shemati\}@uidaho.edu.}}
	\markboth{IEEE Transactions on Communications}%
	{H. Mani and S. Hemati, ``Symbolic Stochastic Chase Decoding of Reed-Solomon and BCH Codes''}
	\maketitle
	\begin{abstract}
		This paper proposes the Symbolic-Stochastic Chase Decoding Algorithm (S-SCA) for the Reed-Solomon (RS) and BCH codes. By efficient usage of void space between constellation points for $q$-ary modulations and using soft information at the input of the decoder, the S-SCA is capable of outperforming conventional Symbolic-Chase algorithm (S-CA) with less computational cost. Since the S-SCA starts with the randomized generation of likely test-vectors, it reduces the complexity to polynomial order and also it does not need to find the least reliable symbols to generate test-vectors. Our simulation results show that by increasing the number of test-vectors, the performance of the algorithm can approach the ML bound.  The S-SCA($1K$) provides near $2$ dB gain in comparison with S-CA($1K$) for $(31, 25)$ RS code using $32$-QAM. Furthermore, the algorithm provides near $3$ dB further gain with $1K$ iteration compared with S-CA($65K$) when $(255, 239)$ RS code is used in an AWGN channel. For the Rayleigh fading channel and the same code, the algorithm provides more that $5$ dB gain. Also for $(63, 57)$ BCH codes and $8$-PSK modulation the proposed algorithm provides $3$dB gain with less complexity.
		
		This decoder is Soft-Input Soft-Output (SISO) decoder and is highly attractive in low power applications. Finally, the Symbolic-Search Bitwise-Transmission Stochastic Chase Algorithm (SSBT-SCA) was introduced for RS codes over BPSK transmission that is capable of generating symbolic test-vectors that reduce complexity and mitigate burst errors.
	\end{abstract}
	
	
	\IEEEpeerreviewmaketitle
	\section{Introduction}\label{Sec:Introduction}
	\IEEEPARstart{A}{n} error correction code $\mathcal{C}$ over a $q$-ary finite field $\mathbb{F}_q$ with codeword length $n$, dimension $k$ and minimum distance $d$ can be modeled as a mapping from $\mathbb{F}_q^k$ (information space) to larger space $\mathbb{F}_q^n$ (codeword space). The parameter $d$ is the radius in which the codeword can be recovered \cite{Proakis2008}.  When a small fraction of the $\mathbb{F}_q^n$ space is used for codewords, a large $d$ can be created. Specifically for Reed-Solomon (RS) codes, the code minimum distance is $d = n-k+1$, which provides the largest possible code minimum distance for any linear code with the same input/output length. RS codes and Bose, Chaudhuri, and Hocquenghem (BCH) codes are linear codes suitable for high rate applications due to their large minimum distance. RS codes can be considered as a non-binary form of the BCH codes. In the same length $n$ and approximately same rate, the BCH codes outperform RS codes, but the main reason that RS codes have frequent usage is that they can correct burst errors, which can occur in many applications like data storage devices. These codes are widely used in storage channels, satellite telecommunications (ETS 300 456), space telemetry systems (CCSDS and NTSS), digital video broadcasting (DVB-x) and wireless broadband systems (IEEE 802.x).
	\par The decoding problem is the problem of finding a codeword $C \in \mathbb{F}_q^n$ within a specific distance from a received code $R \in \mathbb{R}_q^n$. The brute-force decoding method suffers from exponential complexity while the code length is increased linearly. The most common decoding algorithm for RS/BCH codes is hard decision decoding Berlekamp-Massey (HDD-BM) algorithm \cite{1054260}. The HDD-BM algorithm has a running time complexity of $\mathcal{O}(n^2)$ and produces unique decoded message, with error correction capability, which is limited by half of the minimum distance, i.e., if number of errors $e \geq \frac{n-k+1}{2}$, then there are several codewords within distance $e$ of the received word and the decoding algorithm cannot find the correct answer \cite{782097,7547499}. 
	\par The HDD algorithms are not able to exploit the potential of error correction capability of these codes. This has motivated researchers to find a better decoding algorithm with decoding capability beyond half of the minimum distance \cite{782097,7547499, Chase,4167665,1661852,Koetter,5557635,212267}. Moreover, in many applications, like concatenated systems, soft output information are required. Consequently, soft-input soft-output (SISO) decoding algorithms for RS/BCH codes are highly attentive. In the following, we briefly go over the main soft decoding algorithms for decoding RS/BCH codes. Lastly, we have to mention that similar to RS codes for BCH codes, HDD-BM algorithm is the simplest decoding algorithm and many of the following soft decoding algorithms can also be applied for decoding of BCH codes.
	\subsection{Algebraic soft decision decoding}
	The main idea behind this method is based on list decoding algorithm, first introduced in \cite{elias1957list,61123}. List decoding algorithm finds all the codewords within specified Hamming distance. In pioneering work by Guruswami and Sudan, the authors proposed a new geometric based list decoding algorithm \cite{782097}. They proved that list decoding problem for RS codes can be modeled by a curve-fitting problem over $\mathbb{F}_q^n$. They presented decoding algorithm with polynomial-time complexity that corrects up to $\floor{n-\sqrt(n.k)-1}$ error which is better that $\frac{n-k+1}{2}$ for any value of $k$. The Guruswami-Sudan (GS) algorithm is a symbol level decoder that contains two major steps: \textit{interpolation} and \textit{factorization} \cite{7547499}. Three years later, Koetter-Vardy (KV), proposed a soft-decision list decoding algorithm, that inherited the algebraic decoding nature of GS algorithm while assigned weighted multiplicities to the symbols. KV algorithm used soft information and significantly outperformed GS algorithm \cite{Koetter}. 
	\par The interpolation part is complicated and makes the complexity of the soft decoding algorithm orders of magnitude higher than that of the HDD algorithm. Many researchers focused to reduce the complexity of the interpolation algorithms. Specifically, in \cite{4557643}, by utilizing the output polynomials constructed by HDD-BM algorithm, authors formulated a new list decoding algorithm as a curve-fitting problem with reduced complexity for RS and BCH codes. The low complexity Chase type (LCC) decoding with reduced complexity factorization (RCF) is presented in \cite{Bellorado2006,5429128}. The LCC with RCF method uses soft information and constructs $2^{\eta}$ test-vectors using $\eta$ least reliable positions. This method reduces complexity by utilizing similarities between test-vectors to avoid extra computation. The complexity of LCC with RCF algorithm increases exponentially by increasing the number of unreliable symbols. Another computational technique to reduce list decoding complexity was re-encoding approach \cite{5695132}. The re-encoding method converts the original interpolation to a reduced interpolation problem, while its complexity is smaller than the original one. More approaches for complexity reduction can be found in \cite{5429128,7247663,7001286,4837873,4684611,6451269}. 
	\par Recently, Chen et al, introduced progressive algebraic Chase decoding (PACD) algorithm that reduces complexity while maintaining the performance\cite{6336761,7547499}. In their proposed method, the test-vectors with higher reliability are decoded before the low reliability test-vectors. Two most reliable values are assigned to $\eta$ least reliable symbols. Since two decisions are made for these $\eta$ symbols, $2^{\eta}$ interpolation test-vectors will be constructed. To improve the performance, the number of unreliable symbols $\eta$, can be increased which exponentially increases the complexity. They also proposed an enhanced-PACD algorithm (E-PACD) that improves the decoding performance without incurring an exponential increase in complexity. Instead of increasing the number of unreliable symbols, the E-PACD algorithm couples PACD with adaptive belief propagation (ADP) decoding. PACD and E-PACD outperform both GS and KV algorithm with less computational cost. 
	\subsection{Iterative soft decoding}
	Iterative decoding and belief propagation (BP) algorithms was originally proposed for low density parity-check (LDPC) codes\cite{1057683}. But they can be applied to other block and convolutional codes\cite{485714} \cite{7355317}. The parity-check matrix of RS code in symbolic form is defined as 
	\[ H = \left( \begin{array}{cccc}
	1 & \beta & \cdots & \beta^{n-1} \\
	1 & \beta^2 & \cdots & \beta^{2(n-1)} \\
	\vdots & \vdots & \ddots & \vdots \\
	1 & \beta^{n-k} & \cdots & \beta^{(n-k)(n-1)} \end{array} \right),\] where $\beta \in \mathbb{F}_q$ is the primitive element \cite{1661852}. The binary expansion of $H$, can be obtained by replacing each element of $H$ by its equivalent $m\times m$ binary sub-matrix. Let $H_b$ denote the binary expansion of the RS code parity-check matrix, which is high density parity-check (HDPC) matrix of size $(N-K)\times N$ where $N = nm$ and $K= km$. Since, HDPC matrices are not sparse and suffer from short cycles, directly applying iterative decoding to HDPC codes cannot provide good BER/FER performance\cite{1294938}. Since, the performance of the decoding depends on the choice of parity-check matrix, in \cite{661116} and\cite{1365296} suitable parity-check matrices based on RS codes were first constructed. 
	\par More recently, an iterative decoding algorithm was proposed that works based on adapting the parity-check matrix (ADP) in each iteration according to the channel reliability \cite{1661852}. Then the bit reliability is updated based on extrinsic reliability vector and damping factor $\alpha$, where $0<\alpha < 1$. Moreover, different variations of the proposed algorithm was presented by the authors to improve the performance and reduce the decoding complexity. The incorporated hard decision decoding outperforms the original ADP. In this algorithm, a direct HDD decoding is applied at the end of each iteration to improve the decoding performance. To guarantee the performance is always better than ADP and HDD the decoding can continue up to a maximum number of iterations to obtain all the codewords and the most likely codeword is selected. 
	\par The layered residual belief propagation (LRBP), which is an improved version of ADP algorithm, was introduced in \cite{7282989}. The Algorithm combined the ADP with informed dynamic scheduling (IDS) to update degree one variable nodes more than one time in each iteration. Similar to ADP, the parity-check matrix is updated at the end of iteration and is followed by hard-decision to further improve the performance.
	\par The iterative soft decoding algorithms based on binary parity-check matrix, can also be applied to BCH codes. Similar to RS codes, directly applying BP algorithm to the corresponding parity-check matrix is not useful because it suffers from cycles of length 4 and is not sparse. Many algorithms tried to find an appropriate equivalent matrix for BP algorithm from original BCH parity matrix. A same approach similar to \cite{1661852} can be used to make adaptive belief propagation algorithm by adapting the parity-check matrix in each iteration\cite{baldi2008low}.
	\subsection{Chase decoding algorithms and reliability based decoding}
	The Chase algorithm (CA) takes advantages of channel soft information, it uses a low complex hard decision decoder as the main block and generates whole set of test-vectors by replacing low reliable bits or symbols.  Then, the decoded codeword will be selected using soft distance between the channel received data and decoded test-vectors. More specifically, in bit-wise Chase algorithm (B-CA), the decoder exhaustively flips up to $\lambda$ least reliable bits and incorporates an algebraic HDD in each step, i.e., B-CA generates all the available combinations $(2^\lambda) $ to find the most likely codeword. Thus the complexity of the B-CA increases exponentially with $\lambda$ and the overall complexity is $\mathcal{O}(n^22^{\lambda})$, for $(n,\,k,\,d)$  RS code. This limits the B-CA to small value of $\lambda$.
	\par In some cases due to burst errors, most of the unreliable bits occurred in one symbol and other errors were scattered on the whole codeword. Now consider the case, where the first $\lambda$ least reliable bits occurred in one symbol, then the B-CA can never recover the correct message. To deal with this case we can generate symbolic test-vectors based on the symbol likelihood. The generalized form of the B-CA is known as Chase-II algorithm, which exploits symbol reliabilities to generate test-vectors. In Chase-II algorithm, for $(n,\,k,\,d)$ RS code with symbols of $m$ bits, $q = 2^m$ different values are assigned to each unreliable symbol. Thus for $\lambda$ least reliable symbols $q^{\lambda}$ test-vectors will be constructed. The S-CA has a running time of complexity of $\mathcal{O}(n^2q^{\lambda})$. In this paper, we name Chase-II algorithm the symbolic Chase Algorithm (S-CA)\cite{Chase,6839376}. 
	\par Both, B-CA and S-CA suffer exponential complexity. For instance, considering S-CA for RS$(255, 239)$, with $\lambda = 2$, i.e., considering the two least-reliable symbols, $256^2 = 2^{16}$ different test-vectors should be considered. This exponential complexity convinced researchers to produce new Chase decoding algorithms with low complexity. Specifically, in \cite{6839376}, authors introduced a two-step reliability information selection for S-CA. Instead of searching for all the combinations for $\lambda$ least reliable symbols, they considered two first least reliable bits in least reliable symbols. Using this two-step reliability selection they significantly improved the decoding time.
	\par The ground breaking idea for reducing the complexity of B-CA while improving the performance was presented in \cite{5557635}. They exploited a stochastic test pattern generation using soft information of the bits. Here we name their proposed methods bit-wise stochastic Chase algorithm (B-SCA). Using the soft information of each bit, they found the probability of each bit to be one. The B-SCA, generates the test-vectors based on a bit-wise stochastic experiment by using the reliability of each bit in the received sequence. Their proposed method does not need the selection of least reliable bits, and its implementation requires just a random generator and comparators. Furthermore, the complexity of the B-SCA is tractable for RS with higher order. Moreover, to improve the performance of the B-SCA the authors introduced two pre-calculation steps. First, they saturated the most reliable bits according to a threshold value $\theta$. This saturation threshold keeps most reliable bits to be flipped during test pattern generation. Besides, they used the optimized Noise Dependent Scaling (NDS) factor called $\beta$, to reduce the average value of reliability. These two steps improves the performance of B-SCA \cite{5557635,1705910}. 
	\par Again, all of these methods can be applied to BCH codes. Specifically, in \cite{Chase_bch} the authors used B-CA and B-SCA for BCH codes, in AWGN channel for BPSK transmission and compared the FER performance with other soft-decision decoders.
	\par In many high data rate applications, error correction codes are combined with higher order modulations. For example in satellite communications both BCH and RS codes are considered with $q$-ary modulations in order to increase the transmission rate. In this paper, we present a new symbolic stochastic Chase algorithm (S-SCA), by focusing on BCH and RS codes, to take advantages of $q$-ary modulations. The proposed algorithm provides a unique decoding scheme that can be applied to any kind of error correction codes. Using S-SCA we are able to use a hidden gain that is achievable from constellation points in $q$-ary modulations. Our simulation results show that the proposed decoding algorithm provides more than $2$ dB gain over HDD and compare favorably with other soft decoding algorithms. When RS codes and BPSK transmission are used the complexity of B-SCA can be reduced, by considering the fact that RS codes can correct symbolic errors. We can generate symbolic test-vectors even while data transmission was in BPSK format. The proposed algorithm is called the symbolic-search bit-wise-transmission stochastic Chase algorithm (SSBT-SCA). In SSBT-SCA the test-vectors are generated in symbolic domain while data transmission was in BPSK format and it works because RS codes are able to correct symbolic errors. The SSBT-SCA decreases the complexity of B-SCA by a factor of $m$ in each iteration, where $m$ is the number of bits in each symbol. Besides, the SSBT-SCA outperforms the B-SCA when burst errors occurred.
	\par The rest of this paper is organized as follows. Section \ref{sec_preliminaries} presents required preliminaries and notations. We demonstrate how to find the reliability of bits and symbols for $q$-ary modulations. Section \ref{sec_fbsca} presents SSBT-SCA, where the conventional B-SCA is speed up by generating the symbolic test-vectors. We assumed BPSK transmission, and show how to find the symbol probability matrix from bit probability vector. Section \ref{sec_ssca} presents our new S-SCA, when $q$-ary modulations are used. The proposed algorithm can be applied to BCH codes and any other binary and non-binary codes. Simulation results are presented in Section \ref{sec_simulation} where, we compare the performance of our new proposed S-SCA for different modulation schemes in both AWGN channel and Rayleigh fading channel. We also apply the S-SCA for both RS codes and BCH codes. Finally we conclude the paper in Section \ref{sec_conclusion}.
\section{Preliminaries}\label{sec_preliminaries}
	\subsection{RS codes and binary BCH codes}
	Let $\mathbb{F}_q = \{0, 1, \alpha, \cdots, \alpha^{q-2}\} = \{0, 1, 2, \cdots, q-1\}$  denotes a $q$-ary finite field, where $\alpha$ is the primitive element and $\mathbb{F}_q$ is an extension field of $\mathbb{F}_2$ for $q = 2^m$, where $m > 0$. Then, an $(n, k, d)$ RS code, can be defined over $\mathbb{F}_q$. This code receives an information block of $k$ symbols denoted by $U = (u_0, u_1, \cdots, u_{k-1})$ and generates a codeword $C = (c_0, c_1, \cdots c_{n-1})$ with $n$ symbols, by adding $n-k$ redundancy symbols. Each specific symbol in information block or codeword block has a binary $m$-bit equivalent vector like $c_i = (x_0, x_2, \cdots, x_{m-1})$. It is useful to denote the message as a polynomial like $U(x)$ of degree $k-1$ with coefficients $\{u_0, u_1, \cdots, u_{k-1}\} \in \mathbb{F}_q$. Then the coding problem is a mapping $\mathcal{C}$ that maps this code to its evaluation at $n$ values of $x$ chosen from $\mathbb{F}_q$. To say in mathematical form, if we write
	\begin{equation}\label{eq_input_polynomial}
	U(x)= \sum_{0}^{k-1} u_i x^i,
	\end{equation}
	then, the code word can be written as:
	\begin{equation}\label{eq_codeword_equation}
	C = (c_0, c_1, \cdots c_{n-1}) = (U(x_0), U(x_1), \cdots, U(x_{n-1})),
	\end{equation}
	where $x_i \in \mathbb{F}_q$ for $0\leq i \leq n-1$. 
	\par The BCH codes are cyclic codes over $\mathbb{F}_p$ that are defined by a parity check matrix over $\mathbb{F}_{p^m}$. In this paper we consider binary BCH codes where the channel alphabets are binary elements and the elements of the parity check matrix are in $\mathbb{F}_{q}$, where $q= 2^m$. For any positive integers $m \geq 3$ and $t\leq 2^{m−1}$, there exists a binary BCH code of length $n = 2^m-1$ and minimum distance $d_{min} \geq 2t+1$, where $t$ is the error correction capability of the code. 
	The generator polynomial $g(x)$ of BCH code is the lowest degree polynomial over $\mathbb{F}_2$ which has $\alpha, \alpha^2, \alpha^3, \cdots, \alpha^{2t}$ as its roots, where $\alpha$ is the primitive element of the extended field $\mathbb{F}_q$. The binary BCH codes and RS codes are specific cases of non-binary BCH codes, and most of the decoding algorithms can be applied to both of them.
	\subsection{Reliability matrix and a-posteriori probability vector}
	After encoding, the codewords can be transmitted over channel with BPSK transmission or $q$-ary modulations like M-ary phase shift keying (M-PSK) and M-ary Quadrature Amplitude Modulations (M-QAM). At the receiver side calculation of soft information depends on the modulation scheme. These soft information can be in probability domain or in log-likelihood ratio (LLR) domain. In the following we introduce the concept of a-posteriori probability (APP) vector and APP matrix, based on the modulation scheme. These two concept are frequently used in the rest of this paper.
	\subsubsection{Binary phase shift keying}\label{app_mat_bpsk}
	The binary phase shift keying (BPSK) modulation maps $0 \rightarrow \sqrt{\epsilon}$ and  $1 \rightarrow -\sqrt{\epsilon}$. Then these signals are transmitted over an AWGN channel with noise variance $\sigma^2$. At the receiver, the soft values can be computed using the noisy received values $Y = (y_0, y_1, \cdots, y_{nm-1})$. In probability domain and in log-likelihood domain these soft information values are:
	\begin{eqnarray}
	\label{eq: probability_of_one} p_i &= &\Pr \{y_i|x_i=1\} = \dfrac{1}{1+exp\left(\dfrac{2y_i}{\sigma^2}\right)},\\
	\label{eq: LLRvalue} r_i &= &\dfrac{2y_i}{\sigma^2}.
	\end{eqnarray}
	\par Using sign of $r_i$, we can estimate bit-wise hard decision of received sequence as : 
	\begin{equation}\label{eq:hardval}
	y_i^{H}= \begin{cases} 1, & \mbox{if } r_i<0 \\ 0, & \mbox{otherwise} \end{cases}.
	\end{equation}
	\par The magnitude of $r_i$, denotes the reliability of received bit in log-likelihood domain. For a binary variable $y_i^{H}$ it is easy to find $\Pr \{y_i|x_i=0\}$ given $p_i = \Pr \{y_i|x_i=1\}$, since
	$\Pr \{y_i|x_i=0\} = 1-p_i$ and so we only need to store one probability value for
	$y_i^{H}$. Then in probability domain each $y_i^{H}$ has equivalent probability value $p_i$ and we are able to denote the bit-wise APP vector by $P = (p_0, p_1, \cdots, p_{mn-1})$. 
	\par We can consider $P$, in symbolic form like $P = (P_0, P_1, \cdots, P_{n-1})$, where each $P_i = (p_{im}, p_{im+1}, \cdots, p_{(i+1)m-1})$ has $m$ specific bit-wise probabilities. Then, each entry $\pi_{ij}$ in the APP matrix $\Pi$, can be defined as
	\begin{equation}\label{eq_entries_Pi_bitwise}
	\pi_{i,j} = Pr\{c_i = j|P_i\}, \mbox{ for } 0\leq i \leq n-1,\mbox{ and } j \in \mathbb{F}_q.
	\end{equation}
	\par Simply, each element $\pi_{ij}$ denotes that with probability $\pi_{ij}$, the $i^{th}$ symbol in the received vector is equal to $j^{th}$ element in $\mathbb{F}_q$. In Section \ref{sec_fbsca}, when we introduce our SSBT-SCA for BPSK modulation we exploit APP matrix $\Pi$ and APP vector $P$ to generate test-vectors and find the best fit. 
	\par The following example shows how the APP matrix $\Pi$ can be found by APP vector $P$. Consider a 3-bit symbols for RS(7, 4) and assume that APP vector is calculated from \eqref{eq: probability_of_one} and given by $P= [$ 0.82, 0.01, 0.92, 0.03, 0.79, 0.99, $\cdots]_{1\times 21}$ . Then, we categorize the probabilities in different blocks each with $m=3$ elements like $P_0 = [0.82, 0.01, 0.92]$, $P_1 = [0.03, 0.79, 0.99]$ and so on. The element $P_i \in P$ stands for $i^{th}$ received symbol. Table \ref{table1}, denotes the first row of the matrix $\Pi$, that is generated from $P_0$. It means that with high probability the first received symbol is $S_1 = (1, 0, 1) = 5$. The other most probable symbols are $(001)_2 = 1$, $(100)_2 = 4$ , .... and $(010)_2 = 2$, respectively. The second row in matrix $\Pi$, can be found using $P_1$ in similar way. By continuing this algorithm for all $7$ received symbols we can generate the probability matrix $\Pi$.
	\begin{table}[!h]
		\caption{First row of Matrix $\Pi$, obtained from $P_0$}
		\begin{center}\label{table1}
			\begin{tabular}{|c|c|c|}
				\hline
				\textbf{Symbols} & \textbf{Binary } & \textbf{Probability}\\
				\textbf{} & \textbf{ Equivalent} & \textbf{Value}\\
				\hline
				0&000&0.18*0.99*0.08 = 0.0143\\
				\hline
				1&001&0.18*0.99*0.92 = 0.1639\\
				\hline
				2&010&0.18*0.01*0.08 = $1.44*10^{-4}$\\
				\hline
				3&011&0.18*0.01*0.92 = 0.0017\\
				\hline
				4&100&  0.82*0.99*0.08 = 0.0649\\
				\hline
				5&101&0.82*0.99*0.92 = 0.7469\\
				\hline
				6&110& 0.82*0.01*0.08 = $6.56*10^{-4}$\\
				\hline
				7&111&0.82*0.01*0.92 = 0.0075\\
				\hline
			\end{tabular}
		\end{center}
	\end{table} 
\newpage
	\subsubsection{q-ary modulation}\label{app_mat_qary}
	\par For $q$-ary modulations, there is a convenient method to find the APP matrix $\Pi$ and APP vector $P$. The ``LLR algorithm'' is the well-known method to find the LLR value for each bit while using $q$-ary modulation. For simplicity we assume that there exists enough bits for $RS(n, k, d)$ to be transmitted by $q$-ary modulation, if not, we can use zero-filling procedure. For simplicity we assume that the modulation order $q$ is equal to $2^m$, this maps each symbol in RS code to specific symbol in constellation point. The general form is straight forward for any $q$-ary modulation and we can use any modulation order using zero-filling.
	\par Assume that $s_j, j \in \{0,1, 2 \cdots q-1\}$ denotes constellation points for $q$-ary modulation. In transmitter, the codeword symbols $C = (c_0, c_1, c_2, \cdots c_{n-1})$ are transmitted by modulator and each symbol $c_i$ is mapped to a single constellation point $s_j$. In addition, let us denote by $Y = (Y_0, Y_1, Y_2, \cdots, Y_{n-1})$ the noisy received symbols under the presence of independent white Gaussian noise with equal variance $\sigma^2$ in both dimensions in the receiver. Simply, for an $m$-bit symbol assuming equal probability for all symbols, the LLR value for an AWGN channel for a bit in the position $b$ is defined as:
	\begin{equation}\label{eq:llr_bit}
	L(b) = \log \left(\dfrac{\sum\limits_{s_j \in \mathcal{S}_0}e^{\left(-\dfrac{||Y_i-s_j||^2}{\sigma^2}\right)}}{\sum\limits_{s_j \in \mathcal{S}_1}e^{\left(-\dfrac{||Y_i-s_j||^2}{\sigma^2}\right)}}\right).
	\end{equation}
	where, $\mathcal{S}_0$ are sets of constellation points with bit $0$, at the given position, and $\mathcal{S}_1$ are sets of constellation points with bit $1$ at the given position. Then the APP vector $P$ can be found using \eqref{eq:llr_bit}.
	\par To find the APP matrix $\Pi$, we can use the same process like BPSK modulation but for $q$-ary modulations there is a more convenient way based on \textit{Euclidean distance}. For any block codes combined with $q$-ary modulation, symbol probabilities can be estimated by calculation of Euclidean distance of each received symbols $Y_i$ from all constellation points $s_j, j \in \{0,1, 2 \cdots q-1 = 2^m-1\}$. The result is a $\Pi_{n*q}$ probability matrix with elements $\pi_{ij}$ equals to
	\begin{equation}\label{eq: P_mat}
	\pi_{ij} = \dfrac{\hat{\pi}_{ij}}{\sum\limits_{j\in GF(2^m)}{\hat{\pi}_{ij}} },
	\end{equation}
	where $\hat{\pi}_{ij}$ is the un-normalized channel-likelihood 
	\begin{equation}
	\hat{\pi}_{ij} = \dfrac{1}{2 \pi \sigma^2 }e^{\left(-\dfrac{||Y_i-s_j||^2}{2\sigma^2}\right)}, 
	\end{equation}
	where $||Y_i-s_j||$ denotes the Euclidean distance between the channel output $Y_i$ and the constellation point $s_j$, where $i \in \{1, 2, \cdots n\}$, $j \in \{1, 2, \cdots 2^m\}$. The element $\pi_{ij}$ in probability matrix $\Pi$ represents that received symbol $Y_i$ with probability $\pi_{ij}$ is equal to constellation point $s_j$. In Section \ref{sec_ssca}, when we introduce our S-SCA for $q$-ary modulations we exploit APP matrix $\Pi$ and APP vector $P$ to generate test-vectors and find the best fit. 
\subsection{Rayleigh Fading Channel}
	The Rayleigh distribution is the best channel model when there is no Line-of-Sight (LOS) path between transmitter and receiver (for example in a heavily built-up urban environment). In such an environment the received signal is the sum of many reflected/scattered waves. Due to the fact that signals travel different paths they will arrive at the receiver with different phases and thus the amplitude of the received signal will experience rapid fluctuation. The multi-path delay spread, leads to time dispersion and frequency selective fading. This is caused when the band width (BW) of the signal is greater than the BW of the channel. In this paper we assume that the fading channel undergoes flat fading and that means that BW of the signal is smaller than BW of the channel \cite{Sklar1997,Proakis2008}. In this environment the most commonly used distribution for the received signal is Rayleigh distribution, expressed as
	\begin{equation}\label{eq:RayleighFading}
	p(r)= \begin{cases} \dfrac{r}{\sigma^2}\exp\left\{-\dfrac{r}{2\sigma^2}\right\}, & \mbox{for } r\geq0 \\ 0, & \mbox{otherwise} \end{cases},
	\end{equation}
	where $r$ denotes the corresponding signal amplitude and $\sigma^2$ is the variance of the in-phase and quadrature phase components. The phase of the complex envelope of the received signal is normally assumed to be uniformly distributed in [-$\pi$,$\pi$]. 
	\par The complex baseband representation of a flat fading channel is 
	\begin{equation}\label{eq:ChannelModelFading}
	y[n] = h[n] x[n] + z[n],
	\end{equation}
	where $\{h[n]\}$ denotes a sequence of fading coefficients and $z[n]$ is independent and identically distributed Gaussian noise with equal variance ($\sigma^2$). For normalization purpose we assumed that $E\left\{\left|h[n]\right|^2\right\} = 1$. We also assume that the transmitter doesn't have any knowledge of the channel realization, but the  receiver has perfect channel information and the channel varies slowly relative to symbol rate. Then the receiver has exact knowledge of the channel status and \eqref{eq:llr_bit} can again be applied to find the LLR values.
\section{Symbolic search Bit-wise Transmission-SCA}\label{sec_fbsca}
	As previously discussed the B-CA uses an exhaustive search with all combinations of unreliable bits and finds the most likely codeword using ML technique. More specifically, B-CA($\tau$) considers $\lambda$ least reliable bits and generates all available combinations $(\tau = 2^\lambda)$, to find the best code. In B-SCA each bit in the test-vector $Y^m = (y_1^m, y_2^m,\cdots, y_i^m \cdots, y_{nm}^m)$ is generated based on reliability of the received bit, i.e., $|p_i-0.5|$, using a random generator. 
	\par When B-SCA is applied for RS codes we can reduce the complexity of the algorithm by generating symbolic test-vectors. While the RS codes correct errors in symbols there is no need to generate bit-wise test-vectors. Instead of generating bit-wise test-vectors we can generate randomized symbolic patterns in a way that most reliable symbols have better chances to be generated. To do this, we have to find the APP matrix $\Pi$ using APP vector $P$, an upfront method was presented in Section \ref{app_mat_bpsk}. Using reliability matrix $\Pi$, we can generate symbolic test-vectors, which reduces the number of test-vectors by a factor of $m$, in each iteration. In addition, when burst errors occurred the SSBT-SCA can outperforms B-SCA.
	\par By defining a threshold $\theta$ for symbol reliabilities, we are able to speed up the algorithm even more. Let $j_i^1 = argmax_{j\in \mathbb{F}_q} \{\pi_{ij}\}$ and $j_i^2 = argmax_{j\in \mathbb{F}_q, j \neq j_i^1} \{\pi_{ij}\}$ denote the column indices of the largest and second largest probability of the row $i$ in $\Pi$. These indices denote the two most probable symbols in the $i^{th}$ position and the value of these symbols are exactly $j_i^1$ and $j_i^2$. Then the following parameter defines the reliability factor for symbol $c_i$ \cite{7547499}
	\begin{equation}\label{eq_reliability_factor}
	\gamma_i = \dfrac{\pi_{i,j_i^2}}{\pi_{i,j_i^1}}.
	\end{equation}
	\par The reliability factor $\gamma_i \in (0, 1)$ is a good test of the reliability of the received symbols. When $\gamma_i$ is close to $1$, it means that the symbol is less reliable and when the value of $\gamma_i$ is close to $0$ it means that the symbol is more reliable. By defining a threshold $\theta$ we are able to prevent the most reliable symbols to be flipped. Simply by sorting all the symbols based on the reliability factor, we can just generate symbols that have lower reliability values. 
	\begin{figure}[!t]
		\begin{center}
			\includegraphics[width=0.52\textwidth,height=0.52\textheight,keepaspectratio]{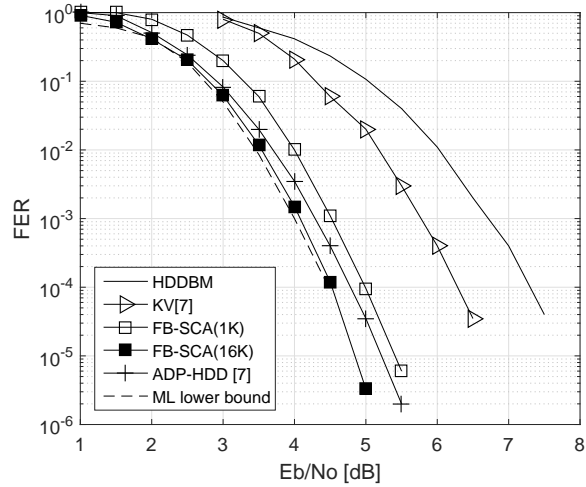}
		\end{center}\caption{The FER performance of the $(31, 25)$ RS code with BPSK transmission over AWGN channel.}\label{fig_rs31_25_bpsk_awgn}
	\end{figure}
	\par
\begin{figure}[!t]
		\begin{center}
			\includegraphics[width=0.5\textwidth,height=0.48\textheight,keepaspectratio]{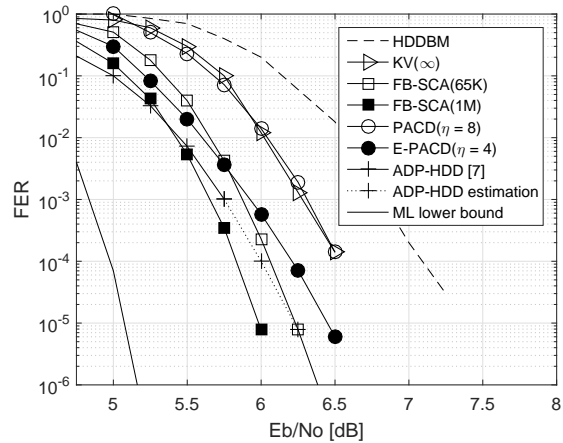} 
		\end{center}\caption{The FER performance of the RS$(255, 239)$ code with BPSK transmission over AWGN channel.}\label{fig_rs_255_239_bpsk_awgn}
	\end{figure}

 Figures \ref{fig_rs31_25_bpsk_awgn} and \ref{fig_rs_255_239_bpsk_awgn} show FER performance of SSBT-SCA when applied to the $(31, 25)$ and $(255, 239)$ RS codes over the additive white Gaussian noise (AWGN) channel, based on BPSK transmission. Our simulation results show that, SSBT-SCA provides the same performance as B-SCA proposed in \cite{5557635}. Due to reduced complexity we are able to increase the number of iterations for SSBT-SCA. We compared the performance of the proposed algorithm with hard-decision decoder as well as soft decoding algorithms, including KV and the ADP-HDD decoder. As denote by KV($\mu$) the Koetter-Vardy algorithm with maximum multiplicity number $\mu$. The PACD and E-PACD algorithms are considered with parameter $\eta$, for $\eta$ unreliable symbols, $2^{\eta}$ interpolation test-vectors are considered. The E-PACD algorithm equipped with an extra adaptive parity-check decoding algorithm. The ADP algorithm is denoted by ADP($A \times A'$). The parameters $A$ and $A'$ denote maximum number of decoding iteration and the number of decoding rounds, respectively. ADP algorithm incorporated with HDD decoding is denoted by ADP($A \times A'$) \& HDD. In the simulations, the parameter $\tau$ stands for the number of iterations for the proposed algorithm.
	\par For the $(31, 25)$ RS codes, the SSBT-SCA with $\tau = 16K$ outperforms the ADP-HDD algorithm and optimal KV performance. Our simulation results show that by increasing $\tau$, the performance of the SSBT-SCA can be improved. In particular, by increasing $\tau$ from $1K$ to $16K$, coding gain of $0.5$ dB can be further achieved for RS$(31, 25)$. Moreover, Fig. \ref{fig_rs31_25_bpsk_awgn} shows that for the $(31, 25)$ RS code, a near maximum-likelihood (ML) performance can be achieved when using the proposed decoding method. However, for the $(255, 239)$ RS code, less significant coding gains are obtained by increasing $\tau$. This is because the $(255, 239)$ RS code has a much larger codebook cardinality. As we can see in Fig. \ref{fig_rs_255_239_bpsk_awgn}, by increasing the number of test-vectors from $2^{16}$ to $2^{20}$, we achieve 0.25 dB gain at FER = $10^{-4}$. This amount is better that any other method with acceptable level of complexity. More specifically, our proposed algorithm outperforms KV and E-PACD algorithms by $0.75$ dB and $0.5$ dB, respectively, and our simulation results indicate that our performance is even better than ADP$(20, 3)$-HDD algorithm.
	In $q$-ary modulations, when symbolic constellation points are transmitted, we can exploit the spatial marginality. The term spatial marginality means that we can generate symbolic test-vectors by considering the reliability of symbols based on their position in constellation. By S-SCA the hidden gain in $q$-ary modulation can be extracted. 
\section{S-SCA}\label{sec_ssca}
	Our proposed symbolic stochastic Chase algorithm (S-SCA) can be applied to any block and convolutional codes concatenated with $q$-ary modulations.  Using $q$-ary modulations, S-SCA provides spatial certainty margin, by generating test-vectors in symbol domain. 
	\par Two factors that prevent the conventional S-CA to be strongly useful algorithm are exponential complexity and brute-force test generation. First, the conventional S-CA($\lambda$) with parameter $\lambda$, generates all the combinations for $\lambda$ least reliable symbols. For $RS(n, k)$ with $q$-ary modulation, the S-CA($\lambda$) requires $q^{\lambda}$ test-vectors to find the most likely codeword. This exponential complexity, limits the performance of S-CA for small value of $\lambda$. Secondly,  without considering the reliability of each symbol it generates all possible combinations for test-vectors. Using symbol reliabilities the brute-force generation of test space is not necessary no longer and test-vectors can be generated in more appropriate way. These two factors drastically reduce the performance of the algorithm. For example, for $RS(n, k)$ the codeword space is $q^n$ while the S-CA searches between $q^{\lambda}$ test-vectors. More specifically, for RS$(255, 239)$, and $256$-QAM modulation, the S-CA($4$) generates $256^4 = 4 Gig$ test-vectors, by considering all possible combinations for four least reliable symbols, which is only $\frac{100}{256^{251}} = 0.000034023*10^{-600}$ percent of the whole code space. The trivial way to increase the number of test-vectors is increasing $\lambda$, which forces exponential complexity.  
	\par Our proposed S-SCA provides an intelligent randomized test-vector generation, which provides the best and most likely set of test-vectors. The S-SCA exploits the spatiality in constellation pattern and generates test-vectors in intelligent manner. Let us denote by $C^t  = (c^t_0, c^t_1, c^t_2, \cdots c^t_{n-1})$ the symbolic test pattern. Each element $c^t_i$ is chosen from the available set of constellation points in $\mathbb{F}_q$ based on the probabilities presented in $i^{th}$ row of the $\Pi_{n \times q}$ matrix. This way symbols with high probability have a better chance of being generated and there is no need to generate all the available combinations. S-SCA reduces the complexity and generates test-vectors that are more likely to be the codeword. 
	\par As discussed in Section \ref{sec_preliminaries}, for $q$-ary modulations the APP matrix $\Pi_{n \times q}$ and APP vector $P$ can be calculated from \eqref{eq: P_mat} and \eqref{eq: probability_of_one}. 
	Again let $C = (c_0, c_1, c_2, \cdots c_{n-1})$ denote symbolic codewords and $Y = (Y_0, Y_1, Y_2, \cdots Y_{n-1})$ denote the noisy received symbols, then using a $q$-ary modulation with constellation points $s_j \in \mathbb{F}_q$ the APP matrix $\Pi_{n \times q}$ can be found from \eqref{eq: P_mat}, where each element of $\Pi_{n \times q}$, can be defined as:
	\begin{equation}\label{eq_entries_Pi_symblwise}
	\pi_{i,j} = Pr\{c_i = s_j|Y_i\}, \mbox{ for } 0\leq i \leq n-1,\mbox{ and } s_j \in \mathbb{F}_q.
	\end{equation}
	\par The \eqref{eq_entries_Pi_symblwise} means that, the element $\pi_{ij} \in \Pi_{n \times q}$, is equal to the probability that $i^{th}$ received symbol is equal to the $j^{th}$ constellation point $s_j \in \mathbb{F}_q$. Again we can use the reliability factor $\gamma_i \in (0, 1)$ from \eqref{eq_reliability_factor} which is the ratio between second and first largest probabilities for $i^{th}$ row in $\Pi_{n \times q}$ matrix. For most reliable symbols the value of $\gamma_i$ is close to 0 and for least reliable symbols the value of $\gamma_i$ is close to 1. 
	\par The generation of the test pattern is straightforward. Let us define an operator $A^s = S(A)$ that sorts each row in a matrix A in ascending order and keeps the permutation indices, to do the inverse action $S^{-1}(.)$. In addition, let us denote the accumulated probability matrix by $AC_{n \times q}$. The $ac_{i,j} \in AC_{n \times q}$ can be found by accumulation of the former elements in $i^{th}$ row of  $\Pi^s_{n \times q}$ up to element $j$. We can interpret elements in $AC_{n \times q}$ as accumulated sub-intervals. Then, the $i^{th}$ test symbol $c_i^t$, can be generated by a uniform random generator and a comparator. We generate a random number $\alpha_i \in (0, 1)$ and find the index of accumulated subinterval for $i^{th}$ row of $AC_{n \times q}$ contains $\alpha_i$. If we note $I_i^j$, the index of subinterval in $i^{th}$ row, then $c_i^t$ can be found using $S^{-1}$ operator, i.e., $c_i^t = S^{-1}(I_i^j)$.
	\par The speed of the algorithm can be improved by simultaneously generating the whole elements of the test-vector in parallel. In addition, by defining a reliability threshold $\theta$, only elements that have reliability less than threshold can be updated. This saturation process provides faster test pattern generation. Furthermore, to improve the performance of the algorithm we can take advantages of NDS method \cite{1705910,5557635}. 
	The proposed algorithm is able to generate a soft information by adding a soft output computation stage at the end of the algorithm. This provides a soft input soft output (SISO) decoder, which is suitable for concatenated codes. The simple pseudo-code for S-SCA is presented in Algorithm \ref{alg:S-SCA}.
	\par The algorithm receives soft channel information and generates soft decoded outputs. In the initialization step, the APP matrix $\Pi_{n \times q}$ and probability vector $P$ can be calculated based on the constellation points and modulation scheme. In the main loop, $\tau$ different symbolic test-vectors are generated. Then, to find the best codeword which has the minimum weighted hamming distance the BM-HDD is applied to the generated test-vector. Simulation results show that S-SCA co-designed with appropriate modulation scheme has better performance in comparison with conventional B-SCA algorithm and S-CA.	
	\begin{small}
		\begin{algorithm}[htp]\label{alg:S-SCA}
			\caption{S-SCA($\tau$)}
			\KwData{$Y = (Y_1, Y_2, \cdots, Y_{n})$ the noisy symbols}
			\KwResult{Decoded data $D$.}
			------------------------------------------------ \;
			\textbf{Initialization : } \\
			\begin{itemize}
				\item Find the APP $\Pi_{n \times q}$ from equation \eqref{eq: P_mat}.
				\item Find the $AC_{n \times q}$ from $\Pi_{n \times q}$.
				\item Find $r_i$ the bit-wise LLR using \eqref{eq:llr_bit}.
				\item Find equivalent $p_i$ value for each $r_i$ from \eqref{eq: probability_of_one}.
				\item Find $Y^H = [y_1^H, y_2^H, \cdots y_i^H, \cdots y_{m*n}^H]$ from \eqref{eq:hardval}.
				\item $Y^m = [\,]$, $i = 0$, iter = 0\; 
			\end{itemize}
			------------------------------------------- \;
			\textbf{Main Loop : }\\
			\While{iter $<\, \tau$}{
				\While{$i\leq\,n$ }{ 
					Generate random symbol $c^t_i$\;
					$Y^m_i$ = Binary equivalent of $c^t_i$.\;
					$Y^m = [Y^m\, Y^m_i]$\;
					$i$++\;
				}
				Perform BM-HDD on $Y^m$ to get $X^m$\;
				Compute the soft weight of $Y^H \oplus X^m$ \:
				$W(Y^H \oplus X^m) = \sum\limits_{i=1}^n |p_i-0.5|(y_i^H \oplus x_i^m)$ ()\;
				iter++\;
			}
			Select decided word $D = X^j$ such that\:
			$D = X^j, W(Y^H \oplus X^j) = \min\limits_{m=1}^{\tau} W(Y^H \oplus X^m)$\;
			------------------------------------------ \;
		\end{algorithm}
	\end{small}
\section{Simulation Results}\label{sec_simulation}
	The comparison of the performance of the S-SCA, when applied to RS codes and binary BCH codes under different modulation schemes and channel noises is presented here. For RS codes we consider the $q$-ary modulations, where elements of RS codes are in $\mathbb{F}_q$. For BCH codes, both BPSK and $q$-ary modulations are reflected, separately. We also reveal both the Rayleigh fading channel and AWGN channel. In this part, S-SCA($\tau$), denotes the symbolic stochastic Chase algorithm. The SSBT-SCA($\tau$) stands for symbolic search bit-wise transmission stochastic Chase algorithm and the S-CA($\tau$) indicates the symbolic Chase algorithm and the parameter $\tau$ is equal to the number of test-vectors. The BM-HDD signifies the hard decision Berlekamp-Massey algorithm. The results for AWGN channel and Rayleigh fading channel are presented separately. Usually, for all the simulation results we talk about gains at specific FER equal to $10^{-4}$, except for the cases where the value of FER was directly mentioned. 
	\subsection{AWGN Channel}
\begin{figure}[!t]
		\begin{center}
			\includegraphics[width=0.48\textwidth,height=0.48\textheight,keepaspectratio]{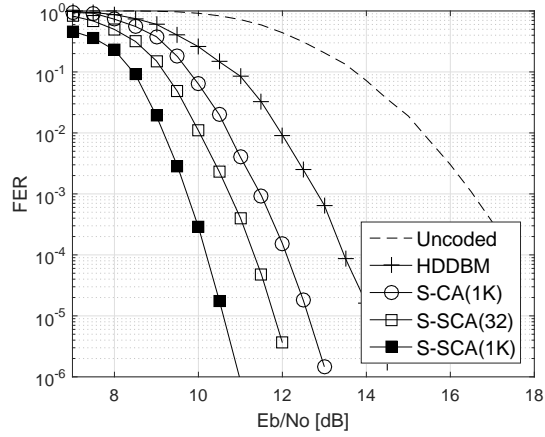} 
		\end{center}\caption{The FER performance of the $(31, 25)$ RS code with 32-QAM modulation over AWGN channel.}\label{fig:m5-SSCA}
	\end{figure}

\begin{figure}[!t]
		\begin{center}
			\includegraphics[width=0.48\textwidth,height=0.48\textheight,keepaspectratio]{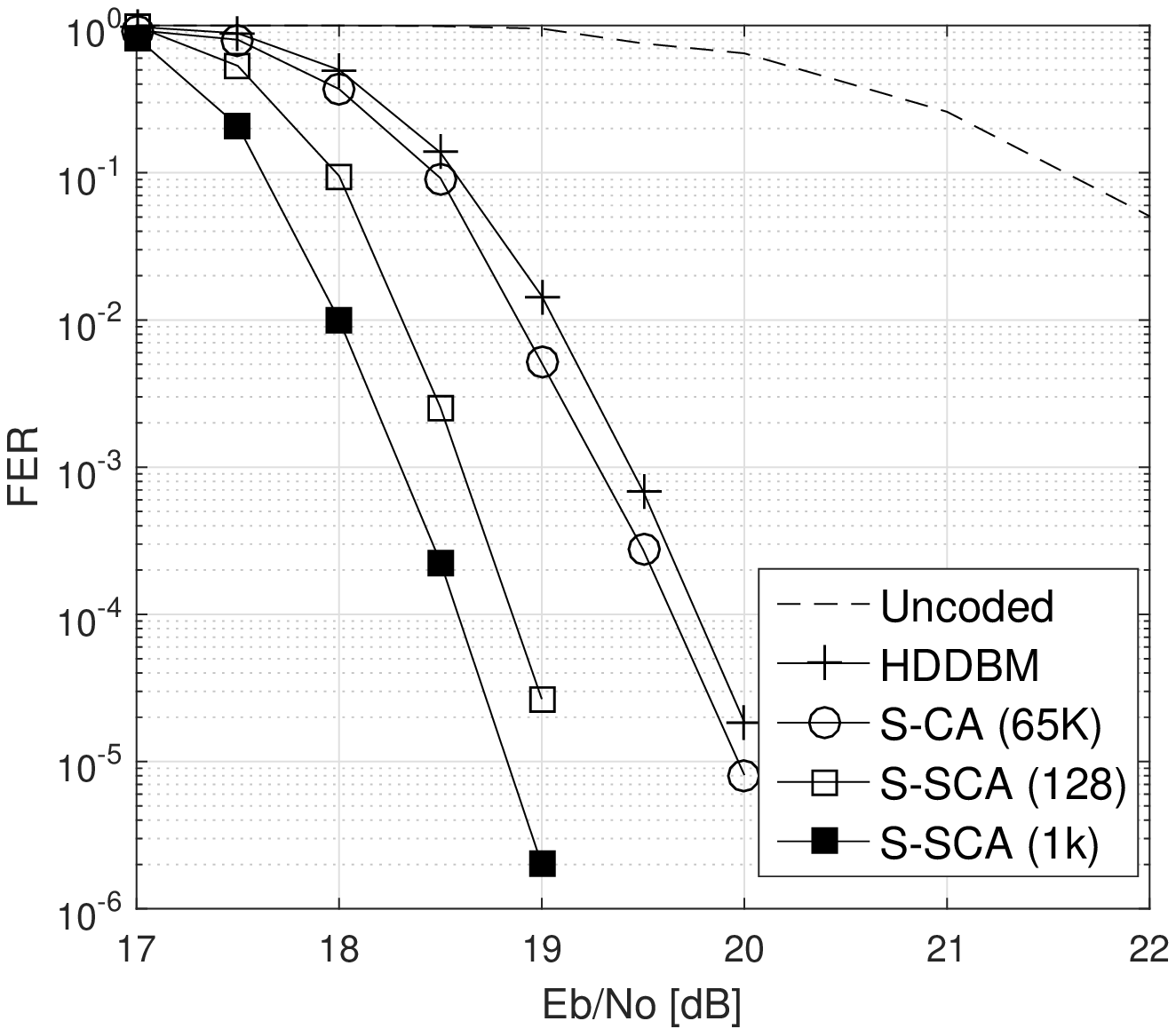}
		\end{center}\caption{The FER performance of the $(255, 239)$ RS code with 256-QAM modulation over AWGN channel.}\label{fig:m8-SSCA}
	\end{figure}

\begin{figure}[!t]
		\begin{center}
			\includegraphics[width=0.48\textwidth,height=0.48\textheight,keepaspectratio]{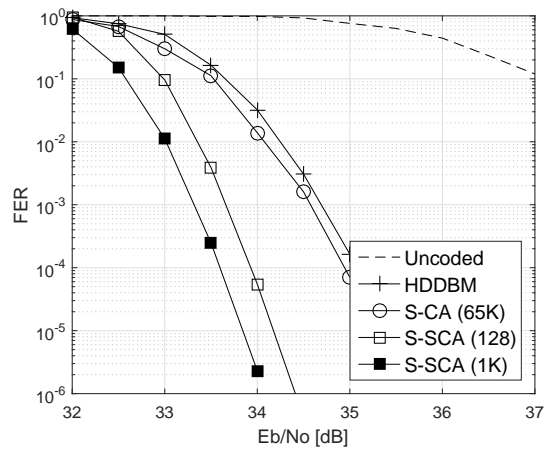}
		\end{center}\caption{The FER performance of the $(255, 239)$ RS code with 256-PSK modulation over AWGN channel.}\label{fig:MPSK_AWGN}
	\end{figure}

\begin{figure}[!t]
		\begin{center}
			\includegraphics[width=0.48\textwidth,height=0.48\textheight,keepaspectratio]{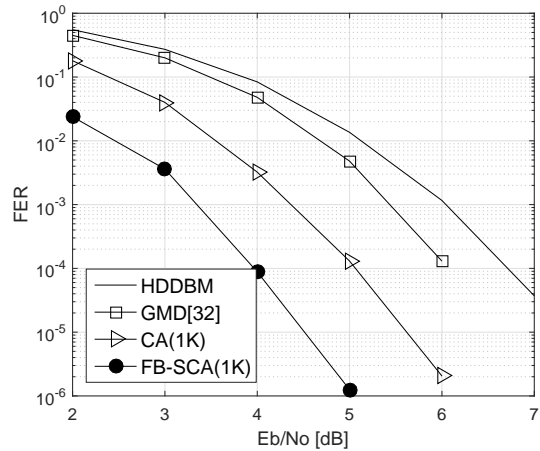}
		\end{center}\caption{The FER performance of the $(63, 30)$ BCH code with BPSK modulation over AWGN channel}\label{fig:bpsk_bch_63_30}
	\end{figure}

\begin{figure}[!t]
		\begin{center}
			\includegraphics[width=0.48\textwidth,height=0.48\textheight,keepaspectratio]{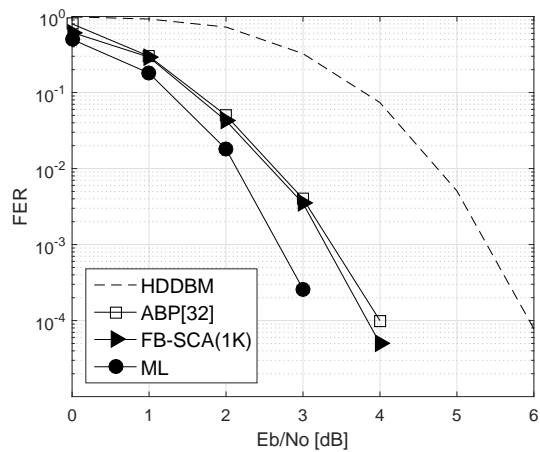}
		\end{center}\caption{The FER performance of the $(127, 71)$ BCH code with BPSK modulation over AWGN channel}\label{fig:bpsk_bch_127_71}
	\end{figure}

\begin{figure}[!t]
		\begin{center}
			\includegraphics[width=0.48\textwidth,height=0.48\textheight,keepaspectratio]{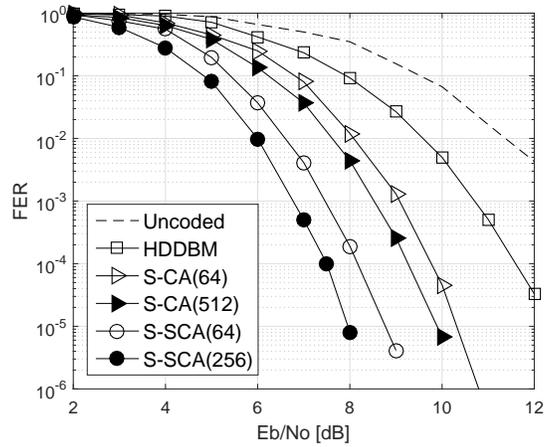}
		\end{center}\caption{The FER performance of the $(63, 57)$ BCH code with $8$-PSK modulation over AWGN channel}\label{fig:8psk_bch_63_57}
	\end{figure}

	Figures \ref{fig:m5-SSCA}-\ref{fig:MPSK_AWGN} show the FER performance of S-SCA when applied to the $(31, 25)$ and $(255, 239)$ RS codes over $q$-ary modulations. Considering HDD algorithm and S-CA, simulation results confirm that S-SCA provides more than $3$ dB gain in $q$-ary modulations. Specifically, Fig. \ref{fig:m8-SSCA} demonstrates that for $256$-QAM modulation and $(255, 239)$ RS code the S-SCA($128$) can outperform the S-CA($65K$). It indicates that the S-SCA requires less than 1\% of test-vectors required by S-CA. Moreover, approximately $0.5$ dB additional gain can be attained by increasing the number of iterations from $128$ to $1K$. Fig. \ref{fig:MPSK_AWGN}, shows simulation results for $256$-PSK modulation. As you can see, the S-SCA($1K$) can outperform the S-CA($65K$) by almost  $1.5$ dB gain.  
	In Figures \ref{fig:bpsk_bch_63_30}-\ref{fig:8psk_bch_63_57}, the proposed algorithm is applied to BCH codes. Figures \ref{fig:bpsk_bch_63_30} and \ref{fig:bpsk_bch_127_71} show the FER performance for SSBT-SCA applied to rate half $(63, 30)$ and $(127, 71)$ BCH codes. The FER performance of the proposed alghorithm is compared with some soft decision decoders. Results show that SSBT-SCA can outperforms all the available soft decoders. Specifically for BCH $(127, 71)$ by $1K$ test-vectors the performance of the algorithm is close to ML bound and also it can outperform the ABP algorithm presented in \cite{baldi2008low}.
	\par The FER performance for high rate BCH code $(63, 57)$ with $8$-PSK modulation is presented in Fig. \ref{fig:8psk_bch_63_57}. The S-SCA with $64$ test-vectors provides approximately $2$dB gain in comparison with S-CA($512$) and it can obtain an additional $0.5$ dB gain by increasing the number of test-vectors to $256$.
	\subsection{Rayleigh Channel}
\begin{figure}[htp]
		\begin{center}
			\includegraphics[width=0.48\textwidth,height=0.48\textheight,keepaspectratio]{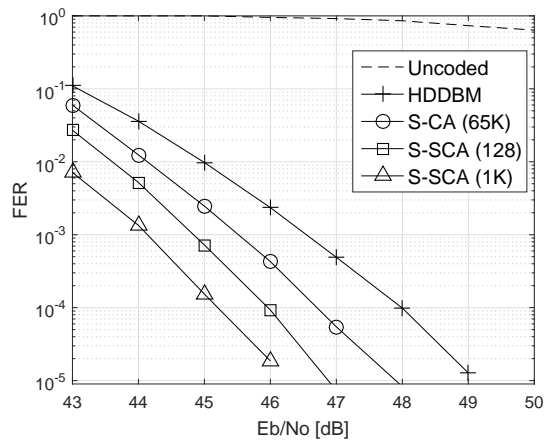}
		\end{center}\caption{The FER performance of the $(255, 239)$ RS code with 256-PSK modulation over Rayleigh fading channel}\label{fig:MPSK_Rayleigh}
	\end{figure}

\begin{figure}[htp]
		\begin{center}
			\includegraphics[width=0.48\textwidth,height=0.48\textheight,keepaspectratio]{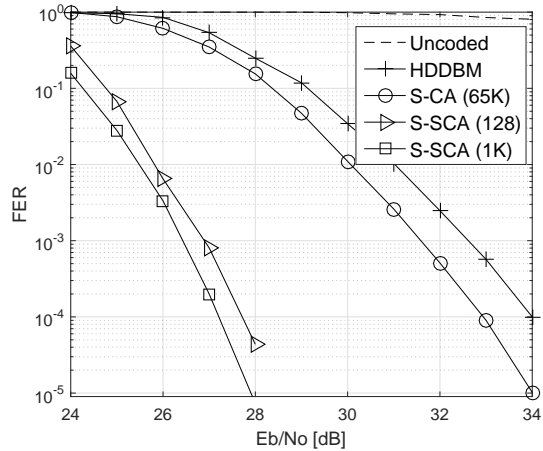}
		\end{center}\caption{The FER performance of the $(255, 239)$ RS code with 256-QAM modulation over Rayleigh fading channel}\label{fig:MQAM_Rayleigh}
	\end{figure}

	The effect of Rayleigh fading channel on the performance of the proposed S-SCA is analyzed in Figures \ref{fig:MPSK_Rayleigh} and \ref{fig:MQAM_Rayleigh}. The simulation results indicate the FER performance for S-SCA compared with other hard and soft decision decoding methods. For both $256$-PSK and $256$-QAM, with $128$ iterations the S-SCA can outperform the S-CA with $16K$ test-vectors. Also, a relaxation factor $\beta$ was used in S-SCA to generate the a-posteriori probability matrix \cite{1705910,5557635}. For all the curves  $100$ frames of error were counted to find the FER probabilities.
\section{Complexity Analysis}
	The HDD-BM algorithm has running time complexity of order $\mathcal{O}(n^2)$, where $n$ is the length of codewords in $\mathbb{F}_q^n$. Considering this fact, the S-CA($\lambda$) which is the mother of S-SCA has the overall complexity of order $\mathcal{O}(q^{\lambda} n^2)$, where parameter $\lambda$ is the number of unreliable symbols in the received sequence in $\mathbb{F}_q^n$. On the other hand, the overall complexity of S-SCA grows polynomially with code length $(n)$ and also the number of iterations $\tau$ at about $\mathcal{O}(\tau n^2)$, where the $\tau$ is not an exponential function of the code parameters. This low complexity of the S-SCA has many other benefits especially from processing energy and system performance perspectives. To see this, we notice that computation time involved in decoding problem for S-CA and the S-SCA are proportional to the number of test-vectors. As it is observed, the number of test-vectors for the S-CA is much larger than that of the S-SCA (see figures in Section \ref{sec_simulation} ).
	\par Moreover, by increasing the number of unreliable symbols, the computation time of the S-CA exponentially grows. As a result in a low power regime or when high performance is necessary, the S-CA may impose an enormous computation load on the processing unit. The longer computation time of the processor means a higher processing energy and higher consumed power. The high processing power will affect the overall system performance. Here, we try to investigate this phenomenon using a simplified model. This model gives us a general idea about such an effect although it may not be a very accurate analysis. First, we assume that the processing power can be considered as the product of the consumed power in the main decoder block, $P_{HDD}$, times the number of iterations. In other words, we have: 
	\begin{equation}
	P_{Tot} = \mbox{Number of iterations} \times P_{HDD}
	\end{equation}
	where $P_{HDD}$ denotes the consumed power in HDD block and $P_{Tot}$ is the total consumed power. The number of iterations can be considered proportional to the order of complexity. Using this fact and by setting $P = 10^{-20}$ Watts, the consumed power for specific performance can be estimated for different decoding methods.
	\begin{figure}[!h]
		\begin{center}
			\includegraphics[width=0.53\textwidth,height=0.58\textheight,keepaspectratio]{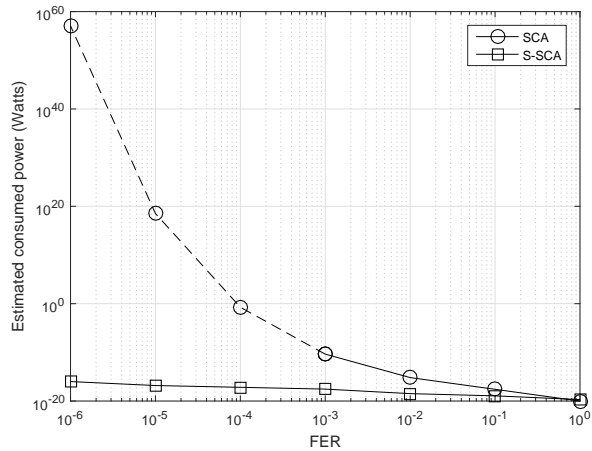}
		\end{center}\caption{The consumed power comparison for decoding of $(255, 239)$ RS code over AWGN channel using $256$-QAM modulation. The consumed energy in HDD block is assumed to be $10^-20$ Watts and the $E_b/N_o = 19 dB$.}\label{fig:consumedPower}
	\end{figure}
	\par As can be seen in Fig. \ref{fig:consumedPower} for the $(255, 239)$ RS code and 256-QAM modulation at $E_b/N_o = 19 $ dB, the S-CA spends a meaningful portion of the available power budget, especially when the low FER is required. However, S-SCA does not require a noticeable processing power. Based on our simulation results for S-CA at $FER = 10^{-3}$ the number of required test-vectors are more than $2^{32}$ and it has exponentially grown. Due to high complexity, the required iteration number was estimated after FER = $10^{-3}$, and we distinguished the simulated part and expected part by robust and dotted lines. Meanwhile, for S-SCA the maximum number of unique test-vector at $FER = 10^{-6}$ is equal to $1K$. The above figure was obtained assuming that $P_{HDD} = 10^{-20}$ Watts and $E_b/N_0 = 19 $ dB. Also based on our simulation results, for the given range of $P_{HDD}$, the consumed processing energy for S-SCA is robust on $P_{HDD}$ while it highly depends on the value of $P_{HDD}$ for the S-CA. 
	\par To determine the complexity of S-SCA, we have to notice that the algorithm looks for the specific individual test-vectors for the decoding process. The trivial way to check the uniqueness of the test vectors is brute-force algorithm, in which each new test-vector compared with all other generated test-vectors. In this case for $\mathcal{O}(N)$ stored test-vector the complexity of finding unique test vector is $\mathcal{O}(N^2)$. The brute-force method suffers from the curse of dimensionality. And the computational costs get higher as the number of test vectors and the length of the code-words $(n)$ increases. Methods like the K-D trees algorithm can compute this at much lower complexity. The K-D tree algorithm uses data sets to build a structured data set named the K-D tree. Then for a newly generated test-vector, it just walked down to the tree until it ends up to one of the regions. Then the algorithm can compare the new test-vector with elements inside the selected area. Building a static K-D tree from $N$ points has the complexity of the order of $\mathcal{O}(N \log N)$. Thus, using the K-D tree algorithm to find the unique test-vectors for the S-SCA, provides the overall complexity, which grows polynomially with code length ($n$) and also the number of iterations $\tau$ to find $N$ unique test-vectors. In other words, the complexity of the S-SCA for $N$ unique test-vector is equal to $\mathcal{O}(n^2\,N \log N )$. As we can see the overall complexity is not a function of unreliable symbols\cite{marsland2015machine}. 
	\par It is clear that the S-SCA can achieve better performance as the number of iteration grows. In other words, the S-SCA asymptotically reaches the ML performance as the number of iteration goes to its maximum limit in $\mathbb{F}_q^n$, which is $q^n$. The significant distinction between the S-CA and S-SCA is the peace of this convergence. Traditionally, the S-CA approaches to ML bound by linearly counting the whole code-word space which, provides exponential complexity. But, in spite of the S-CA, the S-SCA looks for the likely code-words by generating most reliable symbols. In this case, the decoder is looking for true code-word in an $n$-dimensional $q$-ary space by randomizing the symbols at the input of the decoder. And the peace of approaching to ML performance is not constant no longer. The algorithm achieves an acceptable FER performance by looking at high reliable test-vectors, which is the small portion of the whole code-word space. 
	\par
\begin{figure}[htp]
		\begin{center}
			\includegraphics[width=0.48\textwidth,height=0.48\textheight,keepaspectratio]{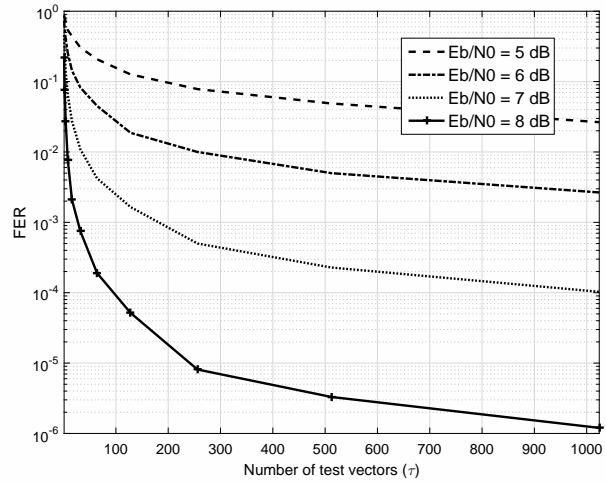}
		\end{center}\caption{The FER performance for decoding of $(63, 57)$ BCH code over AWGN channel using $8$-PSK modulation by increasing the average number of test-vectors}\label{fig:fervsiteration}
	\end{figure}

\begin{figure}[htp]
		\begin{center}
			\includegraphics[width=0.48\textwidth,height=0.48\textheight,keepaspectratio]{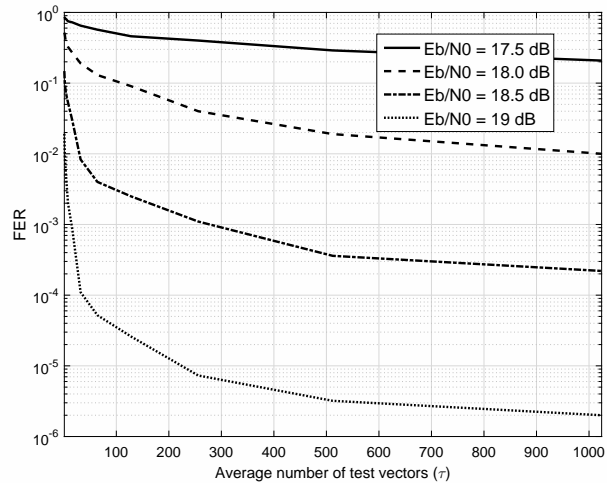}%
		\end{center}\caption{The FER performance for decoding of $(255, 239)$ RS code over AWGN channel using 256-QAM modulation by increasing the average number of test-vectors}\label{fig:fervsiteration_rs}
	\end{figure}

 A precise analysis of the peace of convergence to ML performance is not a straight forward task, especially for $q$-ary modulations. But, there is a trade-off between the acceptable FER performance and the required number of iterations $\tau$ for different channel condition. Here we tried to demonstrate this fact by simulations. In figures. \ref{fig:fervsiteration}-\ref{fig:fervsiteration_rs}, the average number of iterations to achieve the particular FER value for RS and BCH codes under different channel conditions are calculated. For each particular curve, $100$ specific frames of error are counted for each SNR value. And the experiment is repeated $1000$ times to find the average number of iterations for each particular SNR value.
	Finding the optimum iteration number is a function of complexity and performance. It's important to find suitable iteration number to attain the acceptable performance. If $\tau$ has high value, it may provide unnecessary complexity for particular FER performance, on the other hands, small values for $\tau$ may not be able to find the best test-vector. Table \ref{table3} shows the gap to ML bound for $(31, 25)$ RS code at FER $10^{-3}$. Different iteration number provided with the gap to ML bound to denote the peace of convergence. Based on our simulation results, by increasing the number of iterations from $32$ to $1K$, we are achieving approximately $1$ dB gain. But by increasing to $64 K$ the S-SCA provides less than $0.2$ dB extra gain. And by increasing the number of iteration from $512K$ to $1M$, the algorithm provides approximately $0.03$ dB gain. \begin{table}[h]
		\caption{Finding the threshold value of $\tau$, for $(31,25)$ RS code at FER $10^{-3}$.}
		\begin{center}\label{table3}
			\begin{tabular}{|c|c|}
				\hline
				\textbf{Iteration number} & \textbf{Gap to ML bound}\\
				\hline
				$32$&$1.37$ dB\\
				\hline
				$1K$&$0.41$ dB\\
				\hline
				$64K$&$0.27$ dB\\
				\hline
				$512K$&$0.076$ dB\\
				\hline
				$1M$&$0.043$ dB\\
				\hline
			\end{tabular}
		\end{center}
	\end{table}
\section{Conclusion}\label{sec_conclusion}
	This paper has proposed a new SISO decoding algorithm called S-SCA that provides a decoding performance near the ML bound for RS and BCH codes, especially when using $q$-ary modulations. 
	The proposed algorithm reduces the complexity of S-CA by generating a randomized test-vector at the input of the channel decoder. Our simulation results show that the proposed algorithm outperforms the conventional S-CA with less complexity. Still, to reduce the complexity, we employed the K-D tree algorithm to search between the pool of test-vectors to generate individual test-vectors. The algorithm produces the test-vectors based on the likelihood of the symbols.
	Also, the proposed decoder is a SISO block, i.e.,. It can generate and receive the soft information in concatenated codes. By adjusting the iteration number, around particular threshold value, the S-SCA can achieve to acceptable performance with low complexity. Thus, the structure of this decoder makes it suitable for the case where limited transmission power is concerning.
\bibliographystyle{IEEEtranTCOM}
\bibliography{IEEEabrv,Bibliography}
\end{document}